\documentclass[9pt,twocolumn,twoside]{opticajnl}
\journal{opticajournal} 

\setboolean{shortarticle}{false}

\usepackage{lineno}
\usepackage{upgreek}
\usepackage{subcaption}

\title{Programmable Optical Filters Based on Feed-Forward Photonic Meshes}

\author[1,*]{Carson G. Valdez}
\author[1]{Anne R. Kroo}
\author[1,2]{Marek Vlk}
\author[1]{Charles Roques-Carmes}
\author[1]{Shanhui Fan}
\author[1]{David A. B. Miller}
\author[1]{Olav Solgaard}

\affil[1]{Stanford University, Ginzton Laboratory, 348 Via Pueblo Mall, Stanford CA 94305}
\affil[2]{Department of Physics and Technology, UiT The Arctic University of Norway, NO-9037 Tromsø}

\affil[*]{carsongv@stanford.edu}

\begin{abstract}
We demonstrate an integrated photonic circuit based on feed forward photonic meshes that can be programmed and reconfigured to perform arbitrary spectral filter functions. We investigate a subset of the available filter functions, demonstrating that a $N=4$ input triangular mesh with $M=3$ layers may be operated via self-configuration algorithms to filter $M$ arbitrary wavelengths from a given input spectrum. The tunable nature of the architecture enables preconfigured filter functions to be swept in the spectral domain continuously over the free spectral range of the device. This removes any strict requirements between the design parameters of the architecture and the center wavelength of a desired filter function. With this architecture, we experimentally demonstrate arbitrary wavelength rejection filters with contrasts as deep as 40 dB. Further, by intentionally selecting the center wavelengths of each filter function to lie along a wavelength grid defined by $\Delta\lambda = {\lambda_{fsr}}/{N}$ we demonstrate deep wavelength division demultiplexing (DWDM) with inter-channel crosstalk between -25 dB and -40 dB. Unlike typical DWDM systems, in this architecture the center wavelength of each channel is not fixed at fabrication and instead may be swept or reordered arbitrarily. This device demonstrates advantages over typical methods for DWDM, Raman spectroscopy, and correlation spectroscopy as well as other applications. 
\end{abstract}

\setboolean{displaycopyright}{false} 

\begin{document}

\maketitle


\normalsize{}

Spectral filters are ubiquitous components used in nearly every optical application spanning telecommunications \cite{Capmany2006, cameron2018}, environmental sensing \cite{Robinson2005, Fiddler2009}, astrophysics \cite{Tennyson2019, Hearnshaw1986}, and biomedical sensing \cite{Krasnok2018}. Integrated spectral filters such as arrayed waveguide gratings \cite{Munoz2002, Stoll2017} and distributed Bragg reflectors \cite{Wang1974, Palo2023} have long offered a path toward miniaturization. Employing additional local heating elements to shift the spectral band of these devices has provided a degree of tunability \cite{Lee23, Liu2020}; however, this method cannot change the spectral profile of the device.

Traditional fixed-function photonic filters offer high performance but lack flexibility, requiring separate device designs and fabrication runs for each target function. Programmable filters enable a single hardware platform to implement a wide range of spectral responses, enabling further applications including all-optical signal processing \cite{Capmany2024}, correlation spectroscopy \cite{Sinclair97} and wavelength tunability for external cavity lasers \cite{Heim2025}. Programmable spectral filters have been realized utilizing a number of recirculating mesh architectures \cite{Capmany2024, Perez16, Catal2023, Wei2025}. In these systems, light is routed through closed-loop paths that provide spectral selectivity via interference and resonance effects. 

\begin{figure*}[ht]
     \centering
     \begin{subfigure}[b]{\textwidth}
         \centering
         \includegraphics[width=\textwidth]{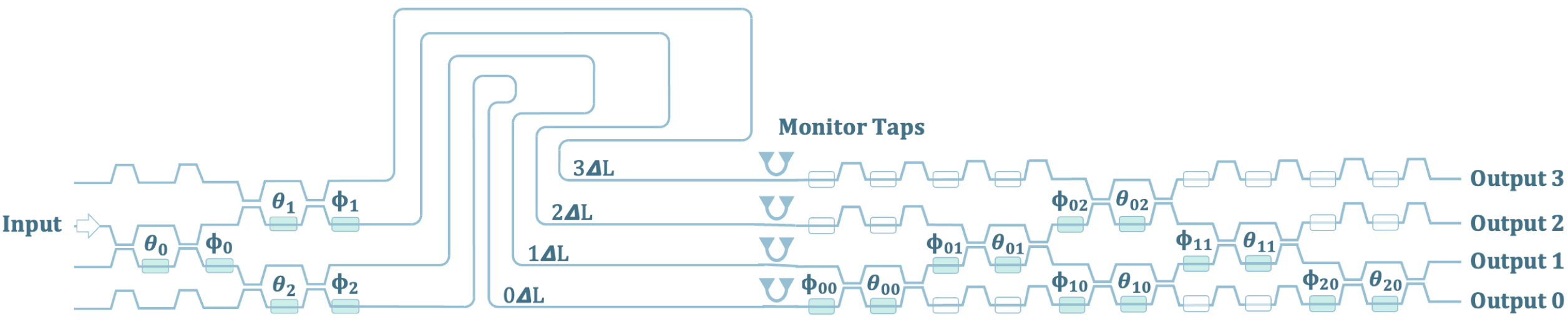}
         \label{fig:schematic}
     \end{subfigure}
     \hfill
     \begin{subfigure}[b]{\textwidth}
         \centering
         \includegraphics[width=\textwidth]{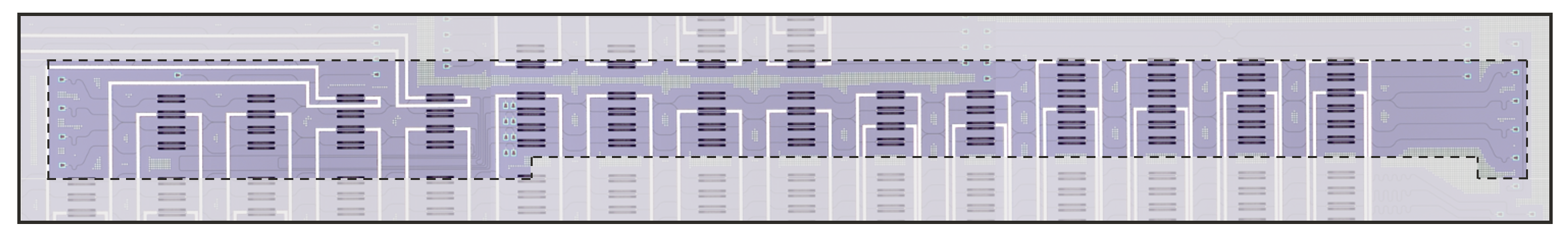}
         \label{fig:image}
     \end{subfigure}
        \caption{(\textbf{Top}) Schematic diagram of a four-channel programmable spectral filter. (\textbf{Bottom}) Microscope image of the fabricated photonic integrated circuit. The photonic circuit is comprised of three subcircuits: a power splitter, an array of waveguide delay lines, and a photonic mesh. Here the power splitting subcircuit has been implemented via a two-stage binary tree of balanced MZIs. The array of $N$ waveguide delay lines introduces the phase shifts $\phi_p$ to each respective waveguide. The array of waveguide delay lines has been designed with a uniform increment $\Delta{L} =$ 740 $\upmu$m to introduce a linear phase tilt. The photonic mesh applies a programmable matrix transformation $O$, described by matrix elements $O_{pq}$. We implement this subcircuit as an $N=4, M=3$ triangular mesh capable of implementing any linear unitary $4\times4$ matrix transformation.}
        \label{fig:4ChannelFilter}
\end{figure*}

Feed-forward based photonic meshes may be operated using progressive optimization algorithms which enable distinct functionalities that have been widely demonstrated for spatial mode processing \cite{Miller13, Miller2013, Miller15, Miller2020, Valdez25}. However, these architectures are typically designed for broadband operation, lacking the strong wavelength dependence necessary to generate spectral filters.

Recently we have laid the analytical groundwork for a photonic integrated circuit (PIC) architecture based on feed-forward photonic meshes capable of performing arbitrary filter functions \cite{miller2025}. Here we experimentally demonstrate that such a device is capable of filtering multiple arbitrary wavelengths while being programmed via a self-configuration algorithm. Such an algorithm automatically adapts a layer of the mesh to reject an arbitrarily given wavelength. Each layer of the photonic mesh is capable of filtering an additional independent wavelength from its respective outputs. This programmable nature of the photonic mesh enables filter functions to be dynamically adjusted post fabrication.

We demonstrate a subset of the array of filter functions which may be implemented by this architecture. From among the possible filter functions, we demonstrate a bandpass filter whose center wavelength may be continuously tuned across a free spectral range. We implement a series of notch filters, demonstrating that multiple arbitrary wavelengths may be rejected from the outputs of the mesh with a contrast as deep as 40 dB. By choosing the filtered wavelengths to be in close proximity to each other, we are able to form a wide band rejection filter spanning a large portion of the free spectral range. Further, by choosing the filtered wavelengths to correspond to a designed wavelength grid, we perform wavelength division demultiplexing. Unlike typical DWDM systems whose wavelength grid is determined at fabrication, the wavelength grid of this architecture may be determined by self-configuration algorithms and adjusted dynamically. 

\vspace{10pt}
\LARGE{\textbf{Results}}

\large{\textbf{Theory and Modeling}} 
\normalsize{}

The architecture of this programmable filter consists of three subcircuits as displayed in figure \ref{fig:4ChannelFilter}: a $1 \times N$ arbitrary power splitter, an array of $N$ waveguide delay lines, and an $M\times{N}$ programmable photonic mesh. In this section, we derive the transfer function of such an architecture, similar to \cite{miller2025}, albeit with a different choice of reference waveguide and indexing system to closely match the physical implementation of the architecture developed in this work. For the analysis of this system, we assert the simplifying assumption that over a reasonable bandwidth, the operations of both the power splitting and photonic mesh subcircuits are wavelength independent. This means that any wavelength dependence of the system results entirely from path length differences introduced in the array of waveguide delay lines. In a practical implementation, this condition is satisfied to first order by ensuring that all possible paths through these subcircuits are of equal length as shown in figure \ref{fig:4ChannelFilter}. 

The power splitting subcircuit divides a single input between $N=4$ output waveguides with programmable splitting ratios such that the output complex fields at each waveguide may be adjusted dynamically. The complex field at each output of the power splitter is given by $a_p$. We assume that the device is lossless such that the input power is conserved as the sum of all output powers. However, a constant overall loss in the circuit may be treated as a pre-factor without impacting the following spectral analysis. This is expressed through Eq. \ref{eq:refname1} where the variable $p$ serves as an index of the $N=4$ output waveguides of the power splitting subcircuit. 

\begin{equation}
\sum_{p=0}^{N-1}|a_p|^2 = 1
\label{eq:refname1}
\end{equation}

The $1\times4$ power-splitting subcircuit has been implemented via a two-stage binary tree of balanced Mach-Zehnder Interferometers, which have internal paths with equal length \cite{Miller2013}. Each MZI is composed of a directional coupler acting as a nominal $50:50$ beam splitter, a thermo-optic phase shifter $\theta_i$ that controls the splitting ratio of each MZI, a second directional coupler acting as a nominal $50:50$ beam combiner, and another thermo-optic phase shifter $\phi_i$ that controls the relative phase between the two outputs of the MZI. Adjusting power to each of the 6 thermo-optic phase shifters in the binary tree enables arbitrary control of the complex splitting ratios between the $N=4$ outputs of the power splitting subcircuit \cite{Miller2020}.


The $N=4$ outputs of the power splitting subcircuit serve as inputs to the subsequent array of waveguide delay lines. Each of the delay lines is a single-mode waveguide with an identical cross-section (500 nm $\times$ 220 nm) and distinct lengths $l_p$, given by Eq. \ref{eq:refname2}. 

\begin{equation}
l_p = l_0 + \Delta{l_p} = l_0 + k_p\Delta{l}
\label{eq:refname2}
\end{equation}

Here $l_0$ is the common waveguide length between all delay lines in the array and $k_p$ is a set of $N$ integers that multiply a unit length $\Delta{l}$ to determine the length increment between waveguide delays. The value $\Delta{l_p}$ then represents the path length difference between the $p^{th}$ and initial ($p = 0$) waveguide delays.

For a given waveguide geometry and material system, the effective refractive index $n_r = \frac{c\beta(\omega)}{\omega}$ of the propagating mode will have an explicit wavelength dependence, influenced by both material dispersion and waveguide dispersion. We model both components of the dispersion by a single term $\frac{\delta{n_r}}{\delta{\omega}}$. For an angular frequency of interest $\omega = \omega_0 + \Delta\omega$ the effective refractive index is given by the linear approximation Eq. \ref{eq:refname3} where $\omega_0$ is the center frequency and $\Delta\omega$ is a frequency offset term.

\begin{equation}
n_r(\omega) = n_r(w_0) + \Delta\omega\frac{\delta{n_r}}{\delta\omega}
\label{eq:refname3}
\end{equation}

The phase delay of the $p^{th}$ waveguide relative to the initial waveguide ($p=0$) may be expressed in terms of the propagation constant and relative delay length of each waveguide:

\begin{equation}
\Delta\phi_p(\Delta\omega) = \beta(\Delta\omega)\Delta{l_p} =  [n_r(\omega_0)\omega_0+n_g\Delta\omega]\frac{\Delta{l_p}}{c}
\label{eq:refname4}
\end{equation}

where $n_g$ denotes the group index of the waveguide delay lines. Expression \ref{eq:refname4} may be simplified further under the condition that, at a central angular frequency $\omega_0$, the relative phase between each output of the arrayed waveguides is an integer multiple of $2\pi$. This condition constitutes a flat wavefront at the output of the array of waveguide delay lines and may be achieved by designing the unit length $\Delta{l}$ to satisfy Eq. \ref{eq:refname5} for a particular central frequency, where $z_p$ is an integer multiple of $2\pi$. 

\begin{equation}
\frac{n_r(\omega_0)\omega_0\Delta{l_p}}{c} = 2\pi{z_p}
\label{eq:refname5}
\end{equation}

Under these circumstances, the relative phase between delay lines as a function of the wavelength detuning is given by Eq. \ref{eq:refname6}:

\begin{equation}
\Delta\phi_p({\Delta\omega}) = \frac{n_g\Delta\omega\Delta{l_p}}{c}
\label{eq:refname6}
\end{equation}

Similarly to the unit length, we can define a unit time delay for the system based on the group index of the guided mode.

\begin{equation}
\Delta{t} = \frac{n_g\Delta{l}}{c}
\label{eq:refname7}
\end{equation}

By substituting Eq. \ref{eq:refname2} and \ref{eq:refname7} into Eq. \ref{eq:refname6} we can further simplify the relative phase delay between waveguides such that it is a simple function of the unit time delay and angular frequency detuning relative to the nominal center frequency of the design. Note that the common $l_0$ term has been dropped from Eq. \ref{eq:refname2} as the $p=0$ waveguide delay line serves as a reference for all values $\Delta\phi_p$. 

\begin{equation}
\Delta\phi_p(\Delta\omega) = k_p\Delta\omega\Delta{t}
\label{eq:refname8}
\end{equation}

As shown in figure \ref{fig:4ChannelFilter}, the outputs of the waveguide delay lines serve as the direct inputs to the $M\times{N}$ photonic mesh of interconnected Mach-Zehnder Interferometers. In this work, we employ a $N = 4$ input triangular mesh. The four input triangular photonic mesh is implemented via $M = 3$ diagonal lines of balanced MZIs, each diagonal line with one fewer MZI than the previous. It has been well established \cite{Miller13, Pai2019, Bogaerts2020} that by tuning each of the $\theta_{mp}$ and $\phi_{mp}$ phase shifters within the mesh, it may be configured to implement any linear, unitary $4\times4$ matrix operation represented by the matrix operator $O$, with elements $O_{qp}$.

The MZIs used in the photonic mesh are nominally identical to those used in the power-splitting subcircuit. We have chosen to use balanced MZIs in both subcircuits as the nominally equal internal path lengths ensure to first order that their operations are wavelength independent.  To ensure an identical propagation length through every potential path through the mesh, we have implemented dummy MZI bends as well as dummy phase-shifting elements which serve no function other than path-length and propagation loss matching. These additional elements help to ensure that any significant wavelength dependence of the PIC can be attributed to the arrayed waveguide delay lines. 

For a photonic mesh represented by the matrix elements $O_{qp}$, the transfer function to each output of the programmable spectral filter is given by Eq. \ref{eq:refname10} where $x(\Delta\omega)$ is the input spectrum to the PIC and $y_q(\Delta\omega)$ is the spectrum at the $q^{th}$ output. 

\begin{equation}
y_q(\Delta\omega) = [\sum_{p=0}^{N-1}a_pO_{qp}e^{-jk_p\Delta\omega\Delta{t}}]x(\Delta\omega)
\label{eq:refname10}
\end{equation}

This expression shows that the phase term will oscillate periodically as a function of the frequency detuning. From this periodicity, we can define the free spectral range (FSR) of the transfer function, which will be solely dependent on the group index and unit length difference of the waveguide delay lines.

\begin{equation}
\omega_{fsr} = \frac{2\pi}{\Delta{t}} \rightarrow f_{fsr} = \frac{c}{n_g\Delta{l_0}}
\label{eq:refname11}
\end{equation}

Within a single FSR, the transfer function is determined entirely by the reconfigurable power splitting ratios $a_p$, the fixed relative delay lengths between the inputs of the photonic mesh $k_p$, and the reconfigurable matrix representation of the mesh itself $O_{qp}$.

As can be seen in figure \ref{fig:4ChannelFilter}, each diagonal line layer of the photonic mesh has a single output which is directly routed to the output of the PIC. The remaining outputs from each layer form the inputs to the subsequent diagonal line of the mesh. Each layer is capable of performing an independent filter function to the inputs of that layer. This is such that an $N=4$ spectral filter may perform as many as $N-1=3$ independent filter functions.

For an arbitrarily configured transfer function or set of transfer functions, the center frequency of the programmable filter may be dynamically tuned by introducing an additional, wavelength-independent phase profile to the inputs of the photonic mesh. This additional phase front does not rely on the introduction of any new components to the generalized architecture. Instead, the existing phase profile may be augmented using programmable phase-shifting elements either in the first layer of the mesh or in the power splitting subcircuit. By introducing an additional phase profile $e^{-j2\pi\gamma{p}}$, a preconfigured transfer function may be shifted in the frequency domain by $\gamma\omega_{fsr}$ where $\gamma$ is a fraction of the FSR to be shifted. This functionality removes strict requirements on the design of the arrayed waveguide delay lines with respect to a desired center frequency.

\begin{equation}
y_q(\Delta\omega) = [\sum_{p=0}^{N-1}a_pO_{qp}e^{-j(k_p\Delta\omega\Delta{t}\ -2\pi\gamma)}]x(\Delta\omega)
\label{eq:refnamegammaeq}
\end{equation}

\large{\textbf{Design and Implementation}}
\normalsize{}

In this section we further investigate a subset of the general architecture described above by enforcing three conditions. First, we require that the power splitter introduces a uniform power distribution between the $N$ waveguide delay lines. Second, the waveguide delay lines are designed to introduce a linear phase tilt across the output waveguides. Third, the matrix elements of the photonic mesh are determined via a self-configuration algorithm \cite{Miller2013, Miller15, Miller2020}.

Given that we have assumed lossless components, the field strengths output from the power splitting subcircuit are given by: 

\begin{equation}
|a_p| = \frac{1}{\sqrt{N}}
\label{eq:refname12}
\end{equation}

This condition may be satisfied through a calibration procedure for the MZI elements of the power-splitting subcircuit. This calibration procedure relies upon waveguide monitor taps which have been introduced following the array of waveguide delay lines. Each monitor employs a directional coupler designed to tap a small portion ($3\%)$ of the power from a bus waveguide. These directional couplers feed directly into a grating coupler designed to emit the tapped power at an angle near perpendicular to the surface of the PIC. The emitted power from each waveguide tap may be monitored while the parameters of the power-splitting circuit are swept, enabling calibration of the two-stage binary tree.

To begin the calibration procedure, a pilot source is launched into a root waveguide of the binary tree at the center wavelength of interest. With a constant input optical power, the following procedure is used to generate a mapping between input heater power and power splitting ratios:

\begin{enumerate}
\item Heater power to the $\theta_1$ phase shifter is swept while monitoring the power transmitted to the $p=2$ and $p=3$ waveguide taps. 
\item Heater power to the $\theta_2$ phase shifter is swept while monitoring the power transmitted to the $p=0$ and $p=1$ waveguide taps. 
\item Heater power to the $\theta_0$ phase shifter is swept while monitoring the sum of powers transmitted to the $p=0$ and $p=1$ waveguide taps and the sum of powers transmitted to the $p=2$ and $p=3$ waveguide taps. 
\end{enumerate}

\begin{figure}[ht]
\centering
\includegraphics[width=\linewidth]{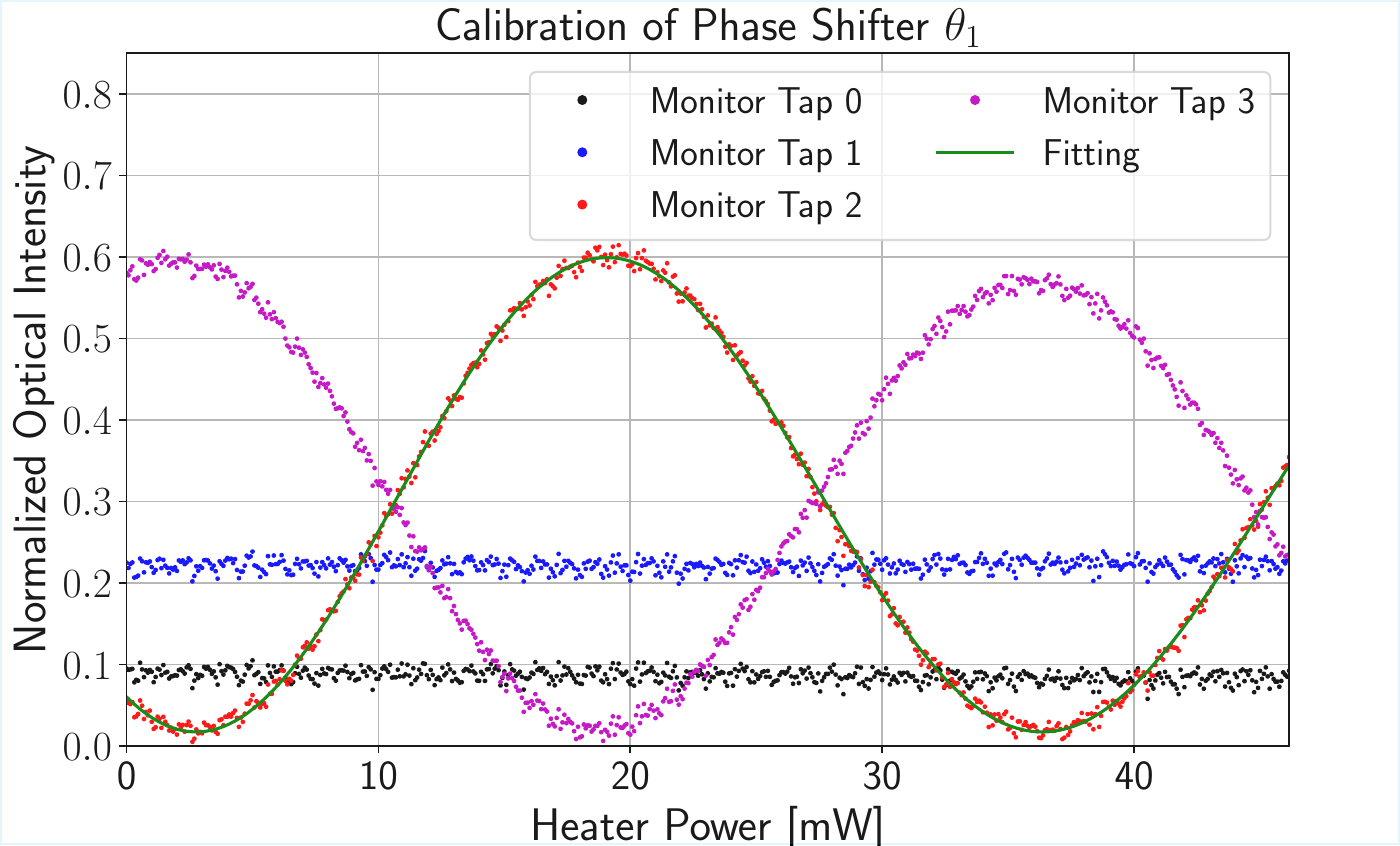}
\caption{Calibration data of the $\theta_1$ phase shifter at a central wavelength $\lambda_0 = $1550 nm. The $\theta_1$ phase shifter controls the split ratio to the $p = 2$ and $p=3$ input waveguides of the photonic mesh and has no impact on the $p = 1$ and $p=0$ input waveguides. We observe the characteristic sinusoidal transmission of a balanced MZI and measure a $P_\pi$ of 18 mW.}
\label{fig:PS_Cal}
\end{figure}

An example of the monitored power through the waveguide taps is given in figure \ref{fig:PS_Cal}. By fitting the monitored power to the characteristic $\sin^2(\theta)$ transmission function of an MZI, we are able to calibrate the heater power versus phase relationship of each thermo-optic phase shifter. Applying this mapping enables the subcircuit to arbitrarily and dynamically adjust splitting ratios. We record a $P_\pi$, the heater power required to drive a $\pi$ phase shift, of 18 mW which is in good agreement with similar thermo-optic phase shifters demonstrated on SOI platforms \cite{Jacques19}.  

To achieve a linear phase front at the output of the array of waveguide delay lines, each consecutive delay line is designed with an identical length increment. 

\begin{equation}
{k_p} = p
\label{eq:refname13}
\end{equation}

The slope of the phase front introduced by these delay lengths is determined by the frequency detuning parameter $\Delta\omega$. At the designed center frequency, $\Delta\omega = 0$, the output of the waveguide delay lines will be a perfectly flat phase profile. For a negative frequency detuning, $\Delta\omega < 0$, the delay lines will introduce a negative phase tilt. Conversely, for a positive frequency detuning, $\Delta\omega > 0$, the delay lines will introduce a positive phase tilt.

The single-mode waveguide used throughout the architecture has a cross-section that is 220 nm tall by 500 nm wide. The fundamental quasi-TE mode has an effective and group index of $n_r = 2.459$ and $n_g=4.056$, respectively, at a central wavelength of 1550 nm. These values have been simulated using the open-source Eigenmode and FDTD solver EMopt.

Based on the modeled group index and Eq. \ref{eq:refname11}, a unit length $\Delta{l} = 740\ \upmu{m}$ has been chosen to achieve an FSR of 100 GHz. This enables a four channel frequency grid with a 25 GHz channel spacing, one of the common International Telecommunication Union (ITU) spacings for use in Dense Wavelength Division Multiplexing. 

The final condition of this subset requires that the matrix elements of the photonic mesh are determined via a self-configuration algorithm \cite{MilerAB2013}. To initialize this algorithm, each element of the mesh is first programmed to the transparent bar state, such that the outputs of the mesh are equal to the inputs. This condition is satisfied when each MZI element of the mesh has been configured to represent an identity matrix:

\begin{figure*}[ht]
     \centering
     \begin{subfigure}[b]{0.49\textwidth}
         \centering
         \includegraphics[width=\textwidth]{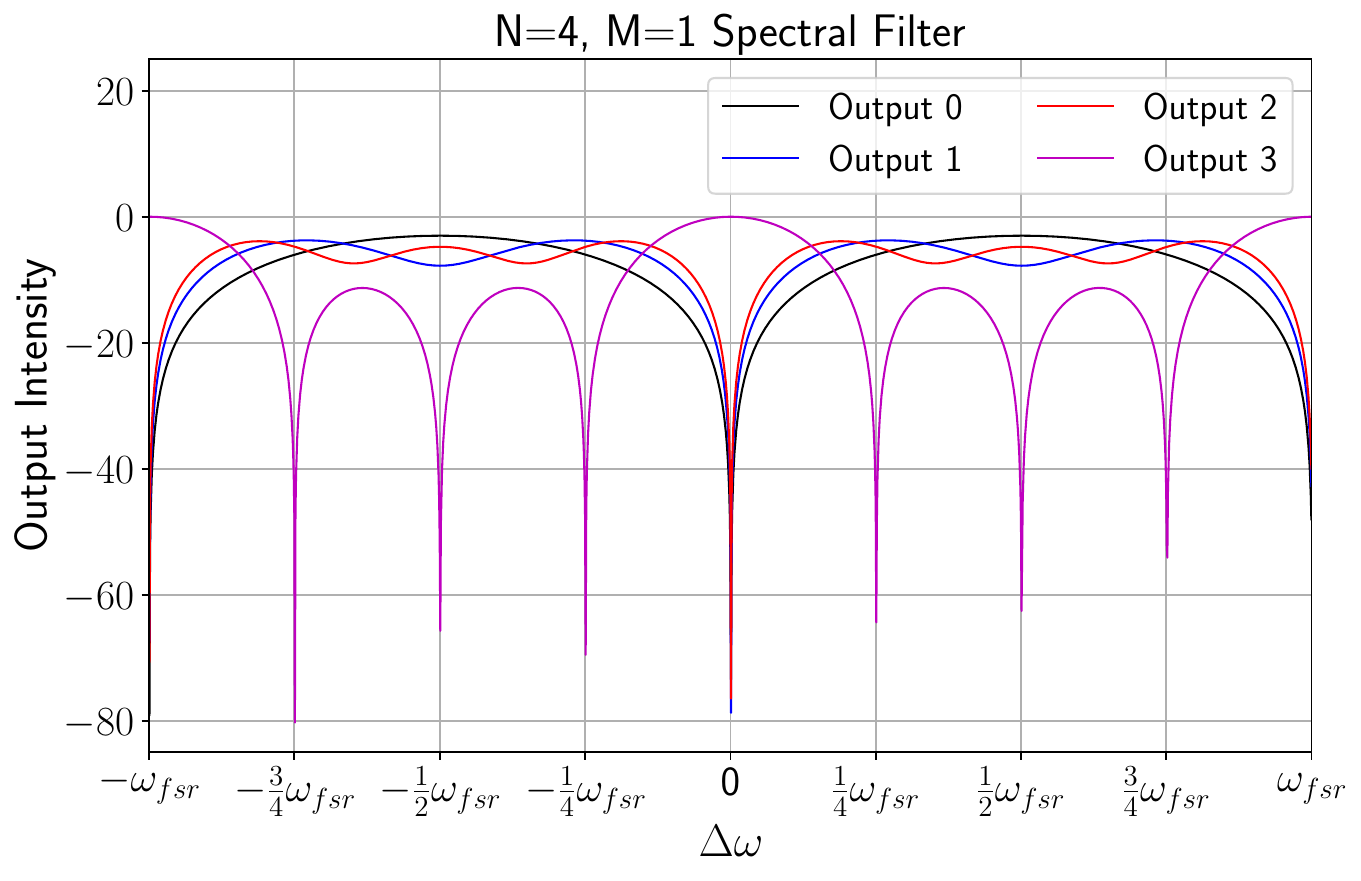}
         \label{fig:y equals x}
     \end{subfigure}
     \hfill
     \begin{subfigure}[b]{0.50\textwidth}
         \centering
         \includegraphics[width=\textwidth]{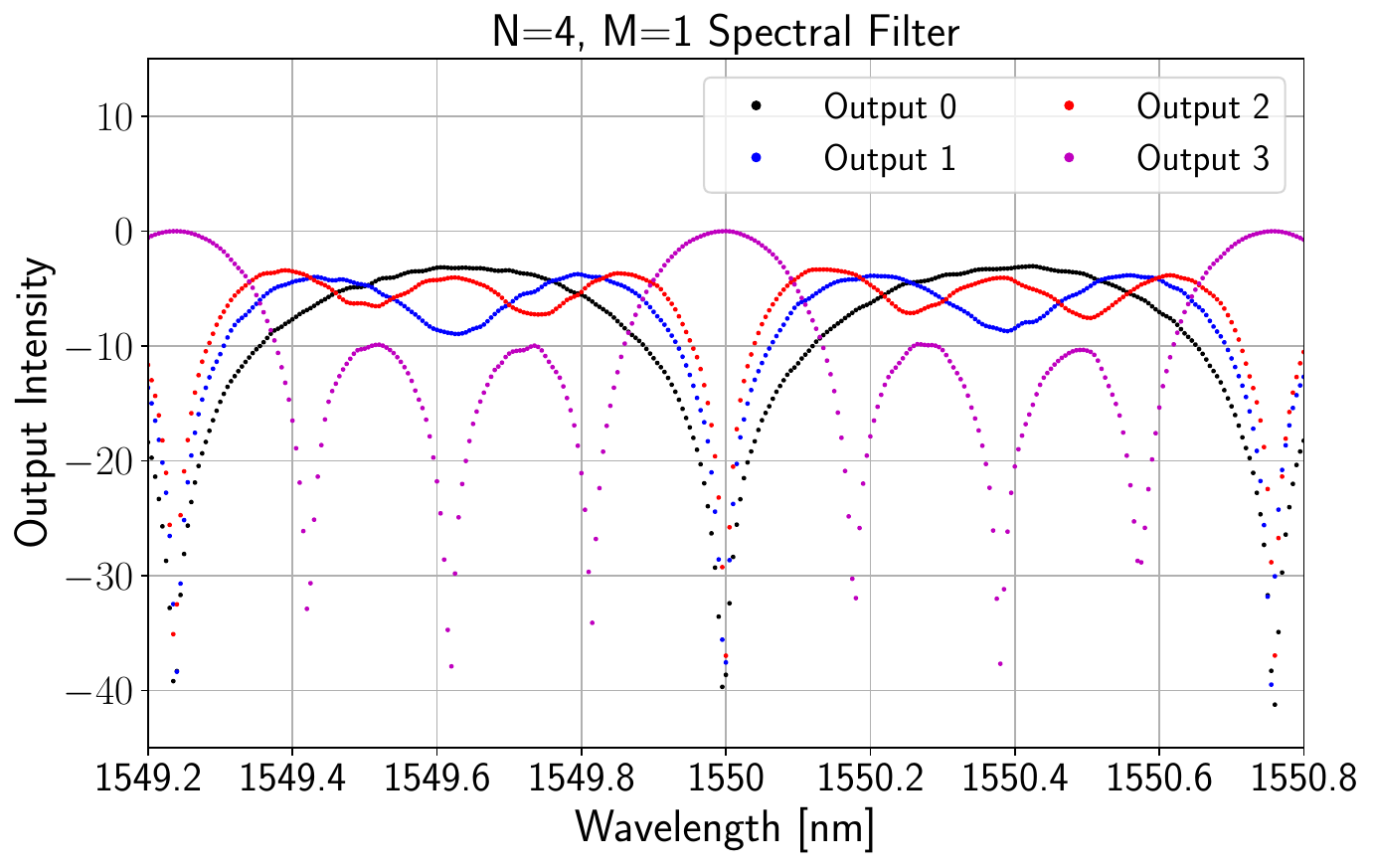}
         \label{fig:five over x}
     \end{subfigure}
        \caption{Simulated (\textbf{Left}) and measured (\textbf{Right}) output spectra of a $N=4$, $M=1$  spectral filter which has been programmed via self-configuration. The output spectra demonstrate that light at the central angular frequency $\omega_0$ corresponding to a central wavelength $\lambda_0 = $ 1550 nm has been collected to the $q=3$ output waveguide and has been completely rejected from the remaining outputs. We measure rejection of the central wavelength from the remaining output channels with over 35 dB of contrast and a free spectral range of 0.78 nm.}
        \label{fig:SingleDiagLine}
\end{figure*}

\begin{equation}
\textrm{Transparent\ Bar\ State}:\ \ \ \ \vec{v} =  \begin{bmatrix} 1 & 0 \\ 0 & 1 \end{bmatrix} \vec{u}
\label{eq:TransBar}
\end{equation}

Initializing each MZI in the mesh to the transparent bar state is necessary for a photonic mesh that operates with only external detectors, as it allows for direct detection of the outputs for each MZI in the mesh. This is critical for the subsequent self-calibration procedure to treat each layer independently. Additional integrated photodetectors may be introduced at the outputs of each MZI to achieve the same effect, making the calibration procedure unnecessary, albeit at the expense of additional circuit complexity. 

The transparent bar states may be achieved through the following progressive calibration algorithm for each of the $\theta_{mp}$ phase shifters in the absence of internal photodetectors as follows \cite{Miller15, Miller2020}: 

\begin{enumerate}
\item Using the calibrated power splitter all input optical power is directed to the $p=0$ input waveguide of the triangular photonic mesh.
\item Heater power to the $\theta_{02}$ phase shifter is swept while monitoring the power transmitted to the $q=3$ output waveguide. 
\item The phase shifter $\theta_{02}$ is set to the transparent cross state which transmits light from the bottom input of the MZI to the top output:

\begin{equation}
\textrm{Transparent\ Cross\ State:}\ \ \ \ \vec{v} =  \begin{bmatrix} 0 & 1 \\ 1 & 0 \end{bmatrix} \vec{u}
\label{eq:TransCross}
\end{equation}

\item Steps 1 through 3 are then repeated for phase shifters $\theta_{01}$ and $\theta_{00}$ progressively completing calibration of the first layer of the mesh.

\item Each element in the $m=0$ diagonal line is set to the transparent bar state, effectively adapting the $N=4, M=3$ triangular mesh to a smaller $N=3, M=2$ triangular mesh.

\item Steps 1 through 4 are iterated for each progressively smaller triangular mesh until every element has been calibrated.
\end{enumerate}

Following completion of the calibration stages, the power splitting subcircuit is set to project a uniform illumination across the $N$ inputs of the photonic mesh. Further, each MZI element of the triangular mesh has been set to the transparent bar state as an initial condition of the self-configuration procedure. To program the filter characteristics, we use a self-configuration algorithm that requires a set of $M$ narrow linewidth pilot sources. Each of the $M$ pilot sources is used to configure the filter function to the corresponding $m^{th}$ layer of the mesh. Assuming that the first pilot source has an angular frequency $\omega_0$, the input vector to the mesh, $\vec{u}_p$, will have a flat phase profile and a uniform intensity distribution as given by expressions \ref{eq:refname5} and \ref{eq:refname12} respectively.

To configure the first layer of the mesh, each MZI is progressively optimized to maximize power out of the top-most waveguide of the associated layer \cite{Miller2013}. This process automatically programs each element to collect all incoming light from the pilot source to the $q = N-1-m$ output waveguide. The output spectrum at the same waveguide may be calculated by applying the conditions of a uniform illumination, a linear phase tilt, and a matrix operator determined via self-configuration to the analytical solution of the general architecture, resulting in Eq. \ref{eq:refname14}  and \ref{eq:refname15}. Here, the phase term $\alpha$ represents the sum of the static and programmable phase terms $\alpha  = \Delta\omega\Delta{t} + 2\pi\gamma$.

\begin{equation}
y(\Delta\omega) = \frac{1}{{N}}[\sum_{p=0}^{N-1}e^{-jp\alpha}]x(\Delta\omega)
\label{eq:refname14}
\end{equation}

\begin{equation}
y(\Delta\omega) = e^{-j\frac{\alpha}{2}(N-1)}\frac{\mathrm{sinc}(N\frac{\alpha}{2})}{\mathrm{sinc}(\frac{\alpha}{2})}x(\Delta\omega)
\label{eq:refname15}
\end{equation}

\large{\textbf{Spectral Filter Functions}}
\normalsize{}

 \begin{figure*}[ht]
     \centering
     \begin{subfigure}[b]{0.483\textwidth}
         \centering
         \includegraphics[width=\textwidth]{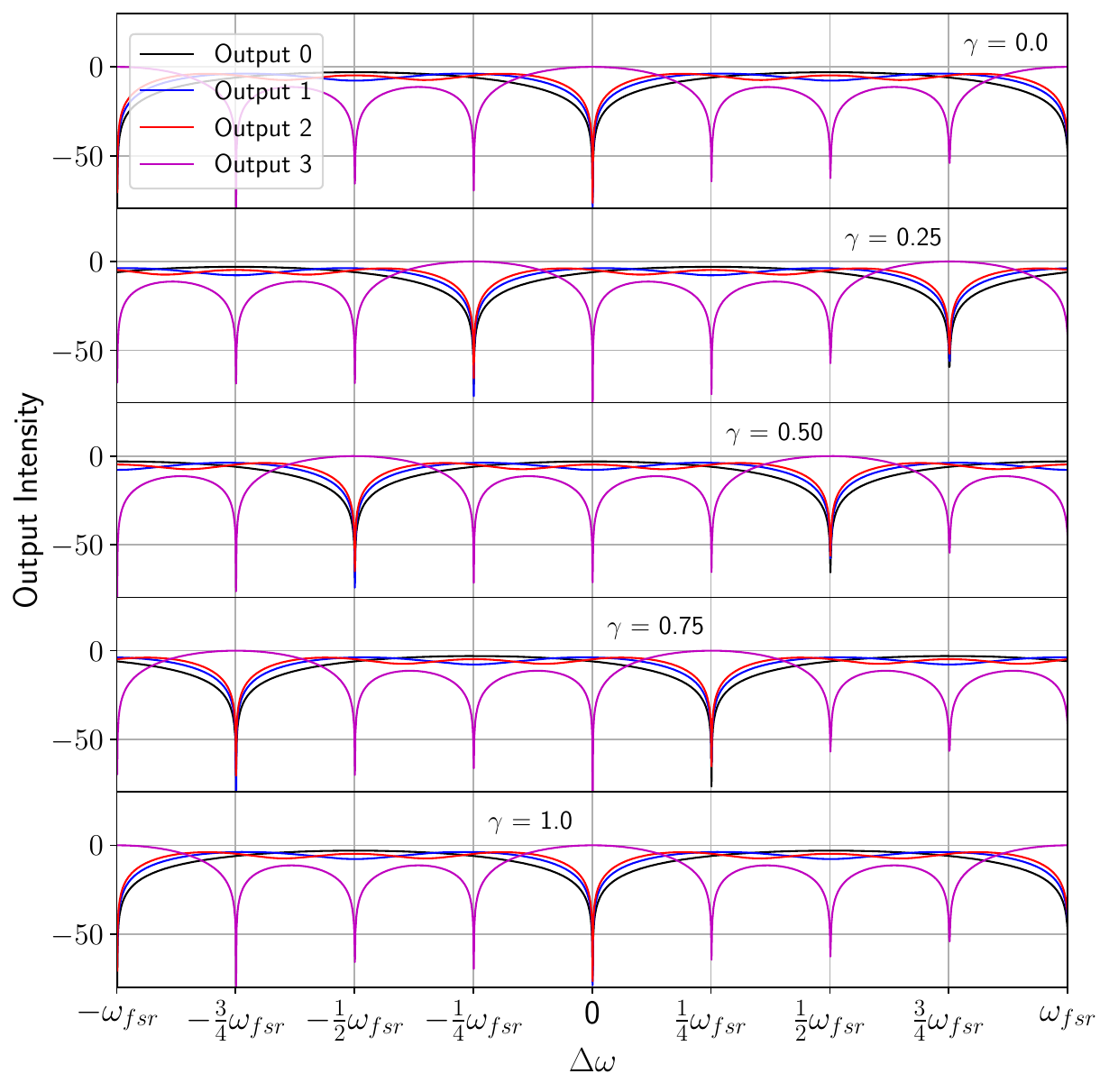}
         \label{fig:GS_sim}
     \end{subfigure}
     \hfill
     \begin{subfigure}[b]{0.48\textwidth}
         \centering
         \includegraphics[width=\textwidth]{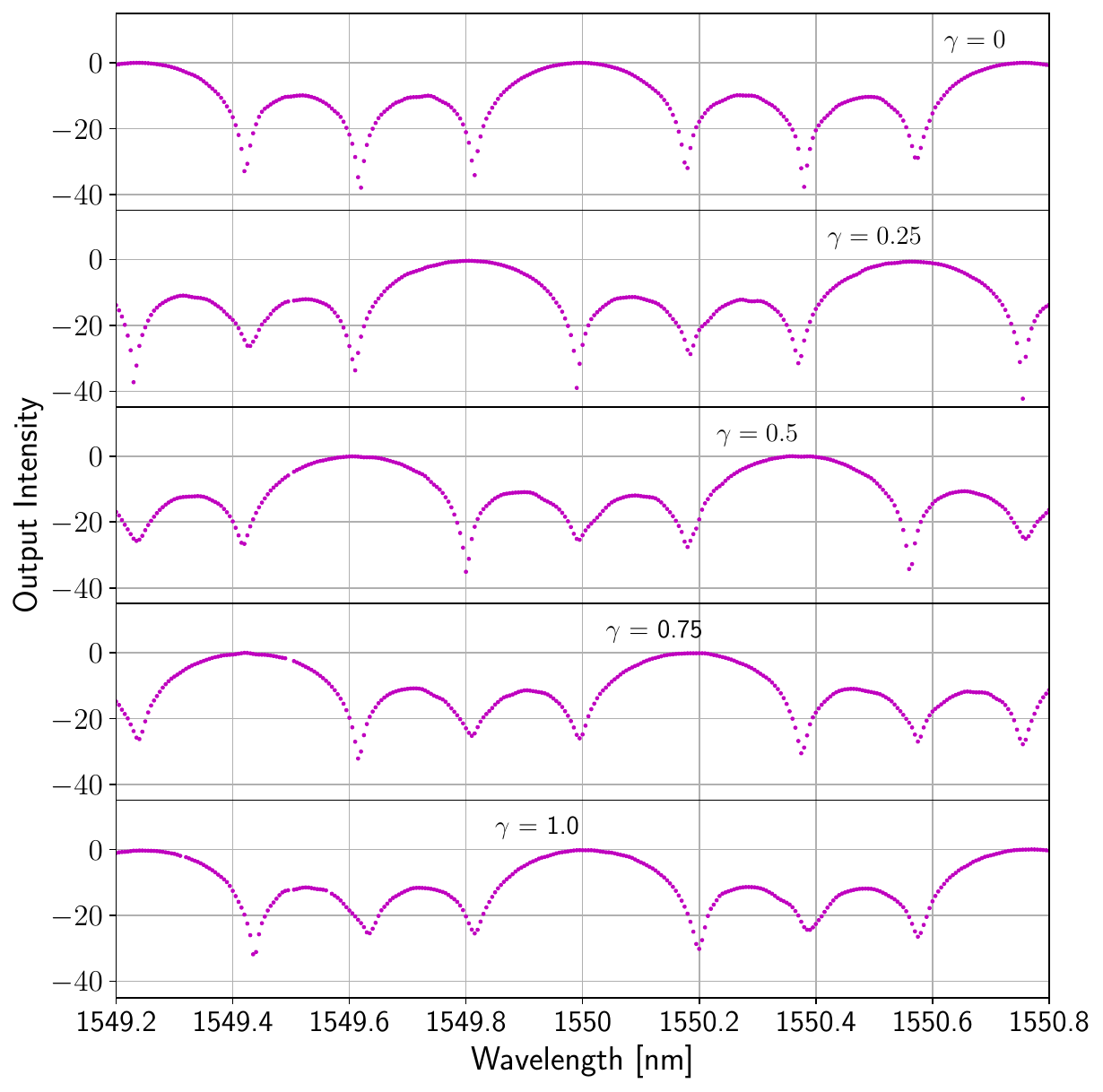}
         \label{fig:GS_exp}
     \end{subfigure}
        \caption{Simulated (\textbf{Left}) and measured (\textbf{Right}) output spectra of a $N=4$, $M=1$  spectral filter while applying an additional phase profile $e^{-j2\pi\gamma{p}}$ for different values of $\gamma$. By adjusting the value of $\gamma$ continuously between 0 and 1, a preconfigured filter function may be tuned continuously across an FSR.}
        \label{fig:Gamma_Shift}
\end{figure*}

The analytical and experimental output spectra given these conditions are plotted in Fig. \ref{fig:SingleDiagLine}. While the analytical expressions in this work are given in terms of angular frequency, by convention the experimental results are expressed in terms of an equivalent wavelength. The single layer filter characterized in Fig. \ref{fig:SingleDiagLine} has been configured from a larger, multi-layer mesh by setting all subsequent layers to the transparent bar states. For this device, we observe that the transmission to the $q=3$ output waveguide contains all of the power associated with the pilot source at $\omega_0$, corresponding to a $\lambda_0 = $ 1550 nm. Consequently, none of the power associated with the pilot source is transmitted to the remaining outputs. We measure that light at the center wavelength has been rejected from the remaining output waveguides with contrasts of 40 dB, 38 dB, and 37 dB for $q=0$, $q=1$, and $q=2$ respectively. 

An experimental FSR of 95 GHz (0.76 nm) is recorded compared to the designed 100 GHz (0.8 nm). We attribute the $5\%$ error to process variability of the single mode waveguide cross section resulting in discrepancies between the modeled and experimental group index. Regardless of such process variations, the filter has been programmatically optimized to filter light of an arbitrary wavelength, determined only by the choice of pilot source.

An additional dynamic, wavelength-independent phase
profile may be introduced using programmable phase
shifters to perturb the static phase profile generated by
the array of waveguide delay lines. This dynamic phase profile may be generated either using the $\phi_{00}$, $\phi_{01}$, and $\phi_{02}$ phase shifters in the $m=0$ layer of the photonic mesh or the $\phi_0$, $\phi_1$, and $\phi_2$ phase shifters of the power splitting sub-circuit. Whether this phase profile is introduced prior to or following the array waveguide delay lines is of no consequence. By programming these phase shifters to introduce a linear tilt in addition to the phase shifts determined by self-configuration, the center wavelength of a preconfigured filter function may be shifted. The programmable linear phase profile is defined by the terms $e^{-j2\pi\gamma{p}}$. By adjusting the value of gamma between 0 and 1, the center wavelength of the preconfigured filter function may be tuned continuously across an FSR. This principle is demonstrated for a $N=4, M=1$ spectral filter for a selection of $\gamma$ values in figure \ref{fig:Gamma_Shift}. The ability to modulate the center wavelength of a preconfigured filter function is a prerequisite for applications such as correlation spectroscopy used in environmental and biomedical sensing \cite{Sinclair97}.

It can be seen in Fig \ref{fig:SingleDiagLine} and \ref{fig:Gamma_Shift} that the output spectrum of the $q=3$ waveguide contains $N-1 = 3$ nulls separated by $\omega_{fsr}/N = $ 23.75 GHz (0.19 nm). Input light at these given frequencies will not be filtered by the first layer of the mesh and will pass through, unattenuated, to the remaining layers. At each of these nulls, light has been rejected from the $q=3$ output waveguide with a contrast between 28 dB and 38 dB. These nulls determine the optimal wavelength grid, where minimal crosstalk between adjacent channels may be achieved. This kind of frequency response is characteristic of arrayed waveguide grating filters, and as such, our device can also emulate such filters. 

It can further be observed from Fig \ref{fig:SingleDiagLine} that outputs $q=1,2,$ and $3$ have multiple transmission peaks with varying relative amplitudes and frequency spacings. The relative amplitude and frequency spacing are functions of the phase shifter settings of the remaining layers of the photonic mesh. These parameters may be adjusted dynamically and arbitrarily, enabling complex functionalities such as intra-FSR pulse shaping. Were this device to act as the wavelength selective component of an external cavity laser, these multiple transmission peaks may be tuned to enable multimode lasing within a single FSR.  

The self-configuration algorithm has effectively generated a spectral filter centered at $\lambda_0 = $ 1550 nm for which all light associated with the pilot source will be routed to the $q=3$ waveguide and rejected from the remaining outputs. Each additional layer of the mesh may generate an additional independent filter function by applying the self-configuration algorithm to progressive layers of the mesh with new pilot sources. This is such that our $N=4$ by $M=3$ spectral filter may reject 3 arbitrary wavelengths from the $q=0$ output waveguide.

 \begin{figure*}[ht]
     \centering
     \begin{subfigure}[b]{0.48\textwidth}
         \centering
         \includegraphics[width=\textwidth]{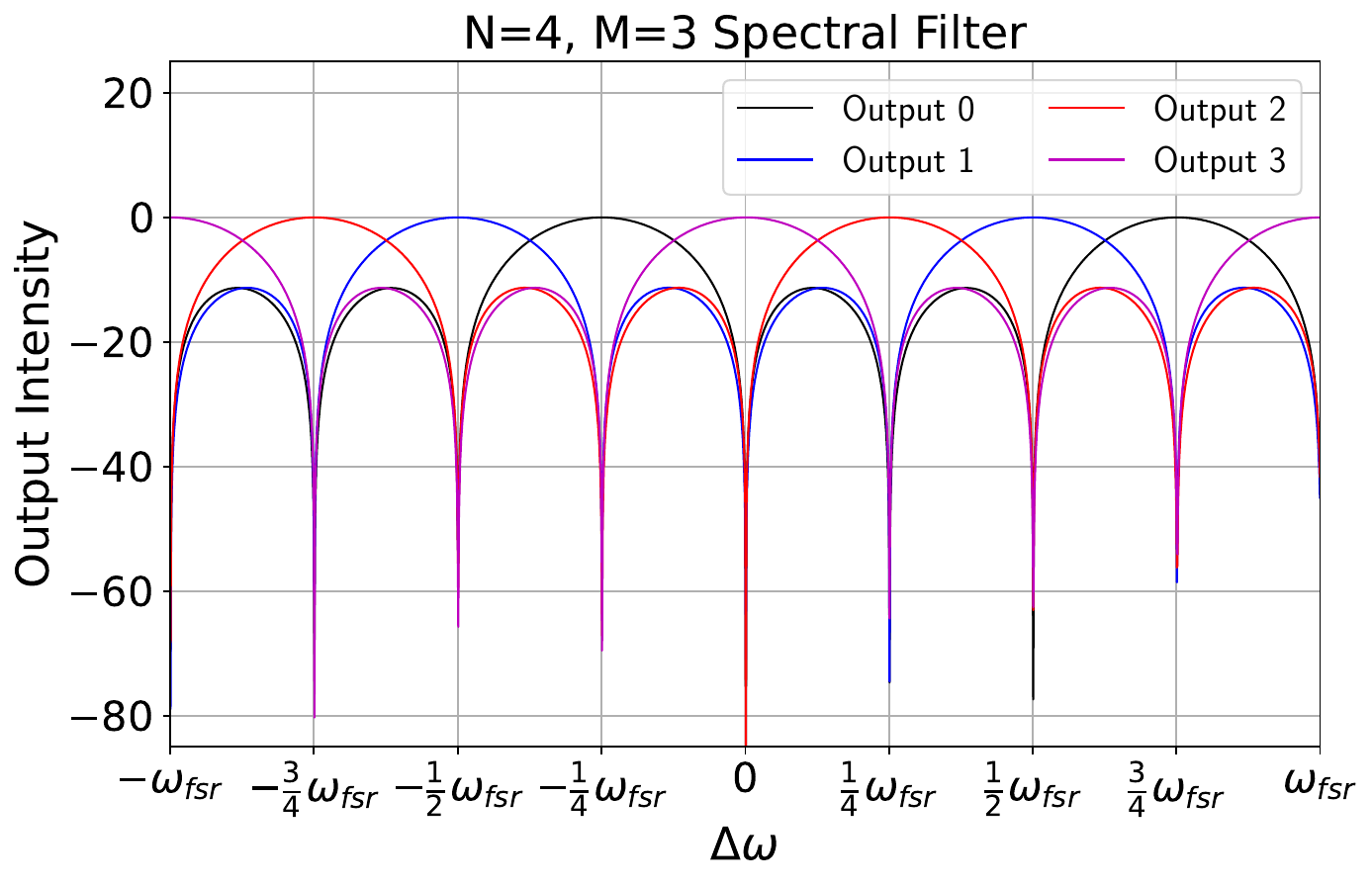}
         \label{fig:4ChannelSim}
     \end{subfigure}
     \hfill
     \begin{subfigure}[b]{0.4875\textwidth}
         \centering
         \includegraphics[width=\textwidth]{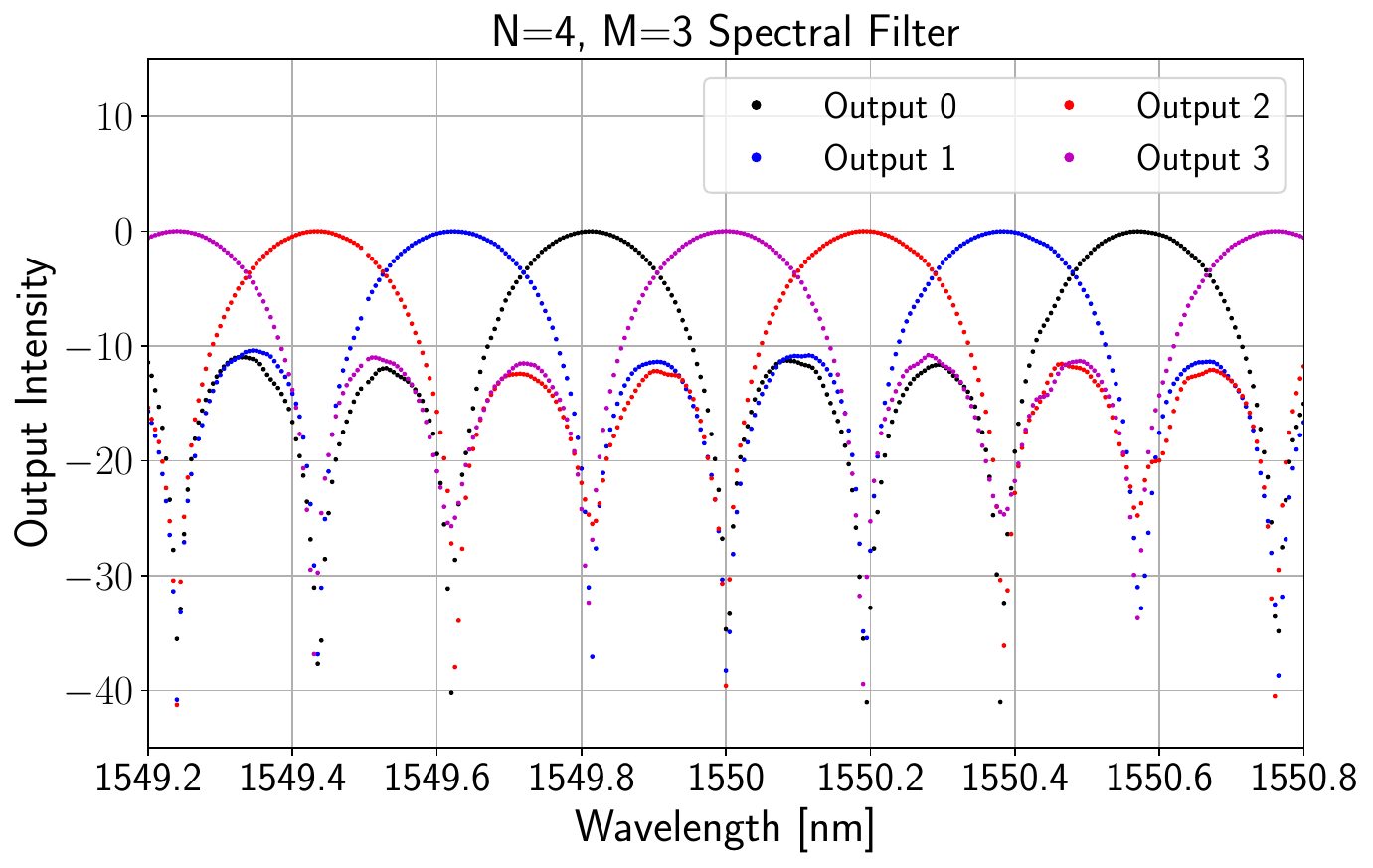}
         \label{fig:4ChannelExp}
     \end{subfigure}
     \hfill
     \begin{subfigure}[b]{0.495\textwidth}
         \centering
         \includegraphics[width=\textwidth]{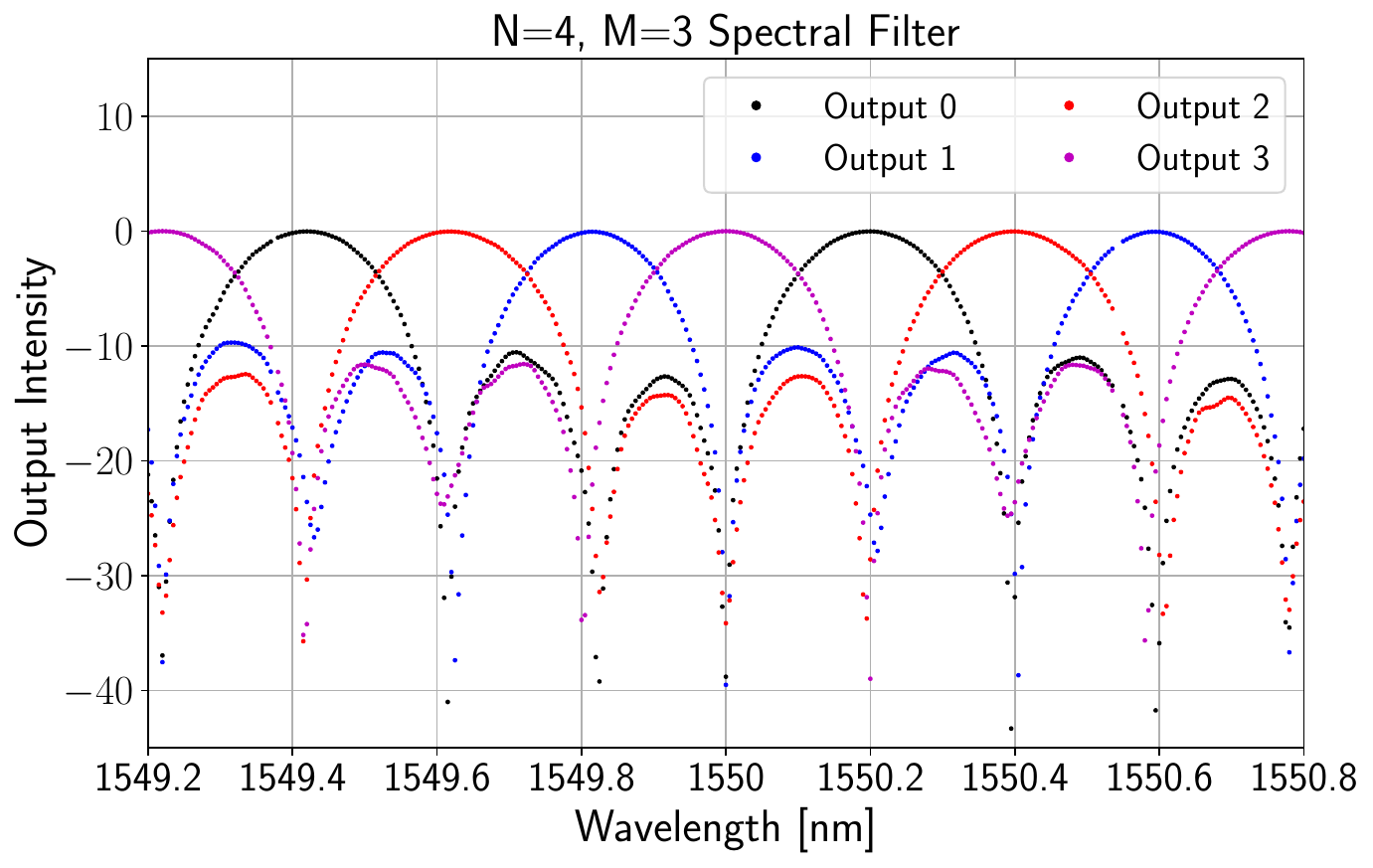}
         \label{fig:SwitchExp}
     \end{subfigure}
     \hfill
     \begin{subfigure}[b]{0.4775\textwidth}
         \centering
         \includegraphics[width=\textwidth]{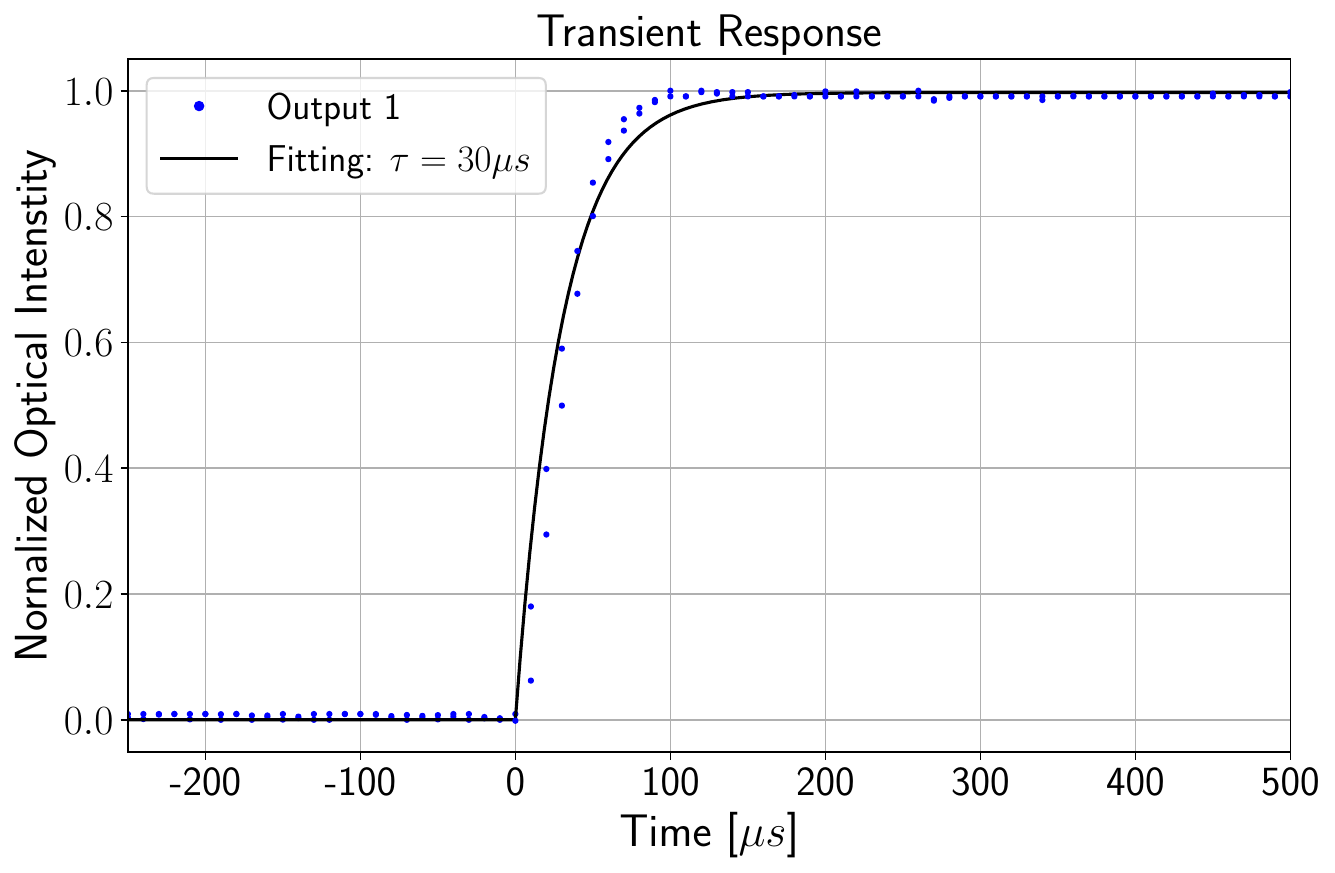}
         \label{fig:SwitchSpeed}
     \end{subfigure}
        \caption{(\textbf{Top Left}) Simulated output spectra of a $N=4$, $M=3$  spectral filter where each layer of the mesh has been programmed via self configuration to reject a frequency associated with the nulls of each other layer of the mesh. The resulting spectra produces high transmission for a single channel of the generated frequency grid while maintaining high rejection each other channel. (\textbf{Top Right}) The chosen wavelengths of interest are $\lambda_0$ = 1550 nm, $\lambda_1$ = 1550.19 nm, and $\lambda_2$ = 1550.38 nm. (\textbf{Bottom Left}) The wavelengths of interest have been reordered here as $\lambda_0$ = 1550 nm, $\lambda_1$ = 1550.38 nm, and $\lambda_2$ = 1550.57 nm to demonstrate that the devices capacity to route any wavelength channel to any output. (\textbf{Bottom Right}) Measured transient response of output waveguide 1 when switching between two preconfigured filter functions.}
        \label{fig:DWDM}
\end{figure*}

By choosing the frequency of subsequent pilot sources to correspond with the nulls of the first layer of the mesh, self-configuration will result in shifted copies of the same transfer function. We demonstrate this principle for a programmable filter with $N=4$ inputs and a complete triangular mesh $M = N-1$. Figure \ref{fig:DWDM} displays the resulting transmission functions for a filter configured to this grid. 

Here the pilot sources have been chosen to have wavelengths corresponding to the nulls at $\lambda_0$ = 1550 nm, $\lambda_1$ = 1550.19 nm, and $\lambda_2$ = 1550.38 nm. Though only three pilot sources have been used to configure this filter, we observe that the remaining output channel exhibits an identical shifted transmission function centered on the remaining null with a corresponding central wavelength of 1550.57 nm. Here each output waveguide achieves high transmission for a single wavelength of interest while maintaining high rejection for the central wavelengths of each other output waveguide. This performs the function of a deep wavelength division demultiplexer, similar to an arrayed waveguide grating filter, and exhibits crosstalk levels between -25 dB and -40 dB. The frequency grid of this device may be arbitrarily shifted across and FSR as given by $\omega_0 + \omega_{fsr}[\gamma + \frac{p}{N}]$. Each output of the filter can be configured to have unity transmission for a single channel on this frequency grid while maintaining high rejection of each other channel. 

In typical DWDM systems, a particular wavelength channel is fixed to a given output waveguide of the device, creating a single wavelength to single output mapping \cite{Capmany2006}. That is not the case for the architecture discussed here. A particular wavelength channel will be coupled to a given output waveguide based solely on the order in which pilot sources are used during the self-configuration process. Such a device is capable of generating an any wavelength to any output mapping on its respective wavelength grid. We demonstrate this principle by repeating the self-configuration procedure to generate a  0.19 nm wavelength grid with wavelengths of interest: $\lambda_0$ = 1550 nm, $\lambda_1$ = 1550.38 nm, and $\lambda_2$ = 1550.57 nm. The resulting filter functions are displayed in Fig. \ref{fig:DWDM}. We observe that similar filter functions have been generated with similar inter-channel crosstalk levels; however, the wavelength bands of interest have been reassigned to new output waveguides. 

 \begin{figure*}[ht]
     \centering
     \begin{subfigure}[b]{0.48\textwidth}
         \centering
         \includegraphics[width=\textwidth]{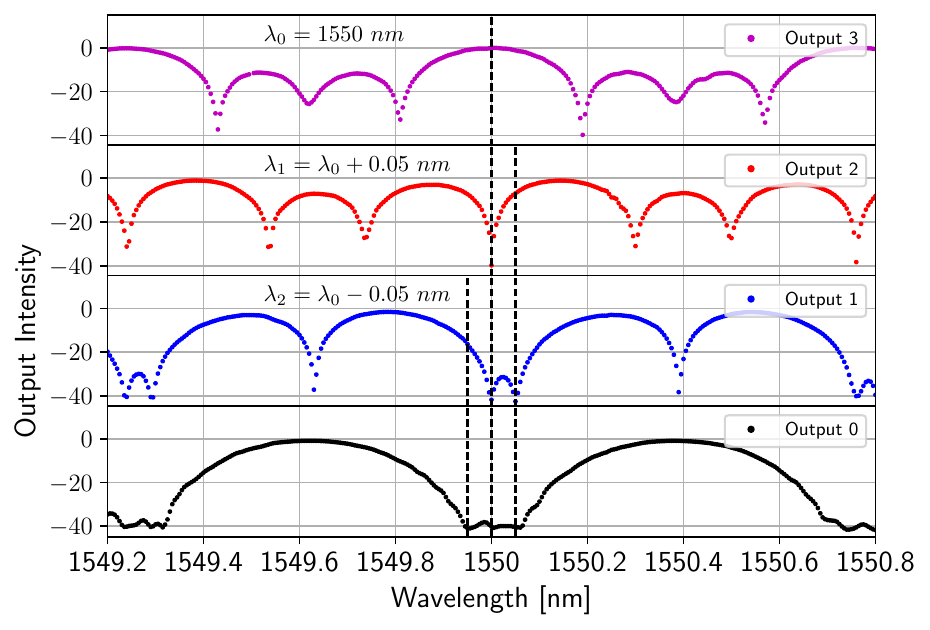}
         \label{fig:Reject3Band}
     \end{subfigure}
     \hfill
     \begin{subfigure}[b]{0.48\textwidth}
         \centering
         \includegraphics[width=\textwidth]{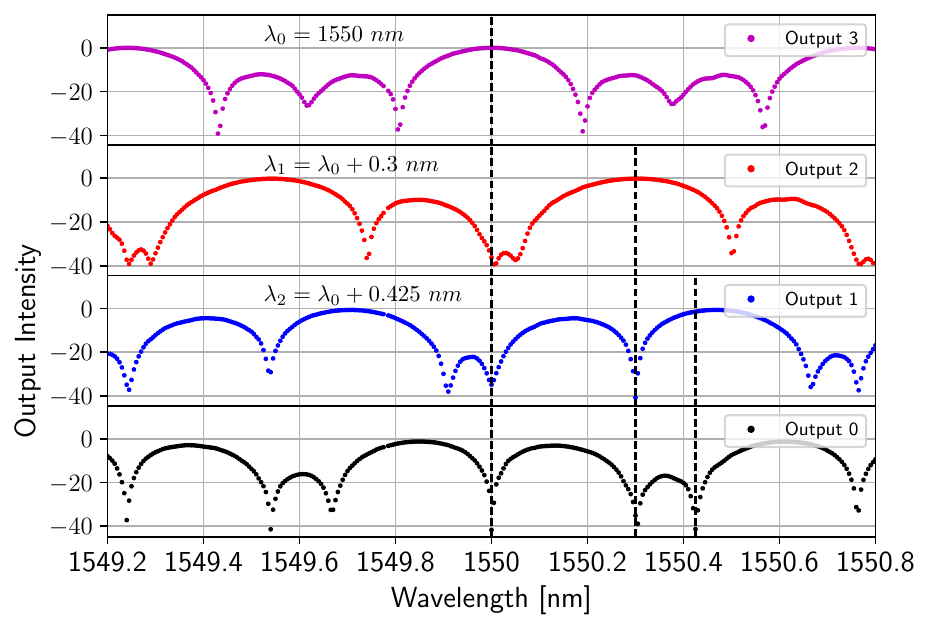}
         \label{fig:Reject3}
     \end{subfigure}
        \caption{(\textbf{Left}) Measured output spectra of a $N=4$, $M=3$  spectral filter where the central wavelength of each filter function has been chosen to be close to each other in the wavelength domain. The corresponding  wavelengths of interest here are $\lambda_0$ = 1550 nm, $\lambda_1$ = 1550.05 nm, and $\lambda_2$ = 1549.95 nm. (\textbf{Right}) Measured output spectra of a $N=4$, $M=3$  spectral filter which has been programmed to apply $M$ filter functions. Here the central wavelength of each filter function has been chosen arbitrarily to be $\lambda_0$ = 1550 nm, $\lambda_1$ = 1550.3 nm, and $\lambda_2$ = 1550.425 nm.}
        \label{fig:Reject}
\end{figure*}

The phase shifter values determined during self-configuration of any given filter functions may be stored digitally and reapplied during future operation of the device. Further, the device can be operated by switching between phase shifter settings for preconfigured filter functions to introduce transient intensity responses at each of the output waveguides. We demonstrate this by  switching between the two preconfigured filter functions which each generate a 0.19 nm wavelength grid. Fig \ref{fig:DWDM} displays the normalized optical intensity measured at output waveguide 1, when a narrow linewidth source with a center wavelength of $\lambda_1 = $ 1550.57 nm has been input to the PIC. Prior to time $t=0$ the filter function rejects $\lambda_1$ strongly at output waveguide 1. At $t = 0$ the phase shifter settings are updated to the wavelength grid which channels $\lambda_1$ to output waveguide 1. The observed transient response takes the shape of a decaying exponential function $y = 1 - e^{-t/\tau}$, with a time constant $\tau$ = 30 $\upmu$s which has been determined by a least squared error fitting method. This time constant corresponds to the thermal time constant of the phase shifters used in this work \cite{Jacques19}, indicating that the switching speed of such a device is limited by the speed of the underlying phase shifting mechanism. 

We further investigate two sub-cases of the programmable filter wherein we have adjusted the center wavelengths of each pilot source.

The first sub-case investigates the spectral response of the filter where each of the pilot sources is chosen to be close together in wavelength such that $|\lambda_m-\lambda_0| << \lambda_{fsr}$. Figure \ref{fig:Reject} displays the results of a filter configured using pilot sources set to 1549.95 nm, 1550.00 nm, and 1550.05 nm. We observe that in this case, rather than each pilot source being discretely filtered, a rejection band forms around the wavelength range of interest. The spectrum of the $q=0$ output waveguide demonstrates a flat rejection band with a contrast of 40 dB over a 0.1 nm bandwidth, which represents ~13$\%$ of the FSR.

The second sub-case allows the choice of center wavelengths for the pilot sources to be chosen arbitrarily over an FSR of the filter. As an example, figure \ref{fig:Reject} displays the spectra from each output waveguide of the filter when the pilot wavelengths have been chosen with no relation. The center wavelengths of each pilot source are chosen to be $\lambda_0 = $ 1550 nm, $\lambda_1 = $ 1550.3 nm, and $\lambda_2 = $ 1550.425 nm. Each filter is configured to reject its corresponding pilot source from transmitting to the remaining output waveguides. This is such that $\lambda_0$ is rejected from the $q=0$, $q=1$, and $q=2$ output waveguides, $\lambda_1$ from the $q=0$ and $q=1$ output waveguides, and $\lambda_2$ from the $q=0$ output waveguide. We observe that each wavelength has been rejected from the corresponding channels with a contrast between 35 dB and 40 dB. In both of these sub-cases, all three wavelength bands associated with the pilot sources used in self-configuration have been well rejected from the $q = 0$ output waveguide regardless of the existing frequency grid. 

\vspace{10pt}
\LARGE{\textbf{Discussion}}
\large{}
\normalsize{}

We have introduced, modeled, and experimentally demonstrated a photonic integrated circuit capable of programmatically performing arbitrary spectral filter functions. The spectral filter may be programmed via self-configuration algorithms that optimize each layer of a photonic mesh to reject an arbitrary spectral line. The transfer function of the device may be designed, optimized, dynamically adjusted, saved, and recalled to accommodate a wide range of applications with a single PIC. 

Among potential applications, this device enables tunable pump wavelength rejection in Raman spectroscopy, continuously tunable bandpass filters for use in external cavity lasers, tunable multimode operation for external cavity lasers, and reprogrammable synthetic spectra generation for use in correlation spectroscopy. The ability to modulate the central wavelength of a preconfigured filter function, enabled by this device, is a prerequisite of filters used in correlation spectroscopy. Further, we have demonstrated the capability of the device to perform DWDM with inter-channel crosstalk between -25 dB and -40 dB. Unlike typical DWDM systems that operate on a static wavelength grid, mapping individual wavelengths sequentially to individual output channels, this device may dynamically shift the wavelength grid and dynamically remap any input wavelength channel to any output waveguides. In essence, this further introduces the functionality of a switching network.

This class of PIC serves as a spectral analog for photonic meshes which have been well researched as spatial mode filters. The flexibility introduced by this architecture enables a wide range of applications spanning the fields of optics and photonics. 

\vspace{10pt}
\LARGE{\textbf{Materials and Methods}}
\large{}
\normalsize{}

Based on the analytical modeling discussed in the previous section, we have designed a programmable spectral filter with a $N=4$ by $M=3$ triangular mesh which has been fabricated by the commercial foundry Advanced Micro Foundry (AMF) Singapore. We have designed this programmable filter for a Silicon-On-Insulator platform with a 220 nm thick silicon device layer, 3 $\upmu$m thick buried silicon dioxide layer, and a silicon dioxide top cladding layer. 

The single-mode waveguide used throughout the architecture has a cross-section that is 220 nm tall by 500 nm wide. The fundamental quasi-TE mode has an effective and group index of $n_r = 2.459$ and $n_g=4.056$, respectively, at a central wavelength of 1550 nm. These values have been simulated using the open-source Eigenmode and FDTD solver EMopt. The coupling of light, both on and off chip, has been managed using surface grating couplers designed to convert between these single-mode silicon waveguides and cleaved single-mode fibers (SMF28). The input and output grating couplers are designed to emit at approximately $12^\circ$ from vertical with approximately 6 dB of insertion loss per coupler.   

We use a HP 81680A tunable laser as an input source to define the pilot frequencies throughout this work. The output of the laser is fiber coupled to a Thorlabs FPC562
manual polarization controller to manage the input polarization state. The output fiber of the polarization controller has been cleaved to ensure good mode matching with surface grating couplers. Both input and output fibers are precisely aligned to their respective grating couplers with a Thorlabs Nanomax 600 series 6-axis stage. 

An over-head observation microscope has been assembled to aid coarse alignment between the input/output fibers and their respective grating couplers. A Xenics Bobcat 640 In-GaAs camera is placed in the detector plane of the microscope, enabling in-situ observation of the incident and output fields. This same microscope assembly enables detection of power at the monitor taps used for calibration of the power splitting subcircuit. Fine alignment between input and output fibers is optimized based on throughput power detected at a HP 81531A photodetector. This same detector is used to measure the output spectra of the programmable filter.

The thermo-optic phase shifters used within each MZI are resistive heaters, patterned in a Titanium Nitride (TiN) layer several microns directly above the waveguiding layer. Each heater is 2 $\upmu$m wide by 120 $\upmu${m} long and has a nominal resistance of 780 $\Omega$. Thermal isolation trenches have been etched through the device layer, buried oxide, and into the silicon substrate on either side of each heater. These trench layers provide a level of mitigation toward thermal crosstalk. Each heating element is independently controlled via a National Instruments digital-to-analog converter (DAC) with 128 analog output channels. 

An interface PCB has been designed to route electrical driving signals from the NI DAC to the PIC. A third-party vendor, Silitronics, has been used for the wire-bonding between the PCB bond pads and the PIC bond pads. During the wire-bonding process, the PIC is permanently fixed to the PCB with a thermally conductive adhesive. For thermal management during the operation of the spectrometer, the assembly is secured to an aluminum mount for heat sinking. Additionally, a pocket has been milled into the aluminum mount for a  thermo-electric cooling (TEC) unit to be inserted directly beneath the PIC.

\begin{backmatter}
\bmsection{Acknowledgments} This work has been funded by  the Air Force Office of Scientific Research (FA9550-21-1-0312, FA9550-23-1-0307), Ames Research Center (80NSSC24M0033),  Marie Skłodowska-Curie Actions (101067268), and Stanford Engineering.

\bmsection{Author Details} $^1$Stanford University, Ginzton Laboratory, 348 Via Pueblo Mall, Stanford CA 94305. $^2$Department of Physics and Technology, UiT The Arctic University of Norway, NO-9037 Tromsø

\bmsection{Author Contributions} D.A.B.M, S.F. and O.S. supervised the project. D.A.B.M, S.F., O.S. M.V., A.K. C.C. and C.V  developed the analytical model. A.K., M.V., and C.V. designed the PIC. C.V and A.K. built the setups. M.V., A.K.,and C.V. performed experimental characterization. C.V wrote the manuscript with input from all authors.

\bmsection{Data availability} Data underlying the results presented in this paper is available from the corresponding authors upon request.

\bmsection{Disclosures} The authors declare no conflicts of interest.

\end{backmatter}

\bibliography{sample}



\ifthenelse{\equal{\journalref}{aop}}{%
\section*{Author Biographies}
\begingroup
\setlength\intextsep{0pt}
\begin{minipage}[t][6.3cm][t]{1.0\textwidth} 
  \begin{wrapfigure}{L}{0.25\textwidth}
    \includegraphics[width=0.25\textwidth]{john_smith.eps}
  \end{wrapfigure}
  \noindent
  {\bfseries John Smith} received his BSc (Mathematics) in 2000 from The University of Maryland. His research interests include lasers and optics.
\end{minipage}
\begin{minipage}{1.0\textwidth}
  \begin{wrapfigure}{L}{0.25\textwidth}
    \includegraphics[width=0.25\textwidth]{alice_smith.eps}
  \end{wrapfigure}
  \noindent
  {\bfseries Alice Smith} also received her BSc (Mathematics) in 2000 from The University of Maryland. Her research interests also include lasers and optics.
\end{minipage}
\endgroup
}{}

\end{document}